\documentclass[conference]{IEEEtran}
\IEEEoverridecommandlockouts
\usepackage{cite}
\usepackage{amsmath,amssymb,amsfonts}
\usepackage{algorithmic}
\usepackage{graphicx}
\usepackage{textcomp}
\usepackage{xcolor}
\def\BibTeX{{\rm B\kern-.05em{\sc i\kern-.025em b}\kern-.08em
    T\kern-.1667em\lower.7ex\hbox{E}\kern-.125emX}}
\usepackage{tabularx}
\usepackage{array}
\usepackage[table]{xcolor}
\usepackage[colorlinks=true, linkcolor=blue, citecolor=blue]{hyperref}
\usepackage{cleveref}
\usepackage{placeins}
\usepackage{xcolor}

\begin{document}
\title{An Ultra-Low-Noise Tunable TIA in 45nm Silicon Photonics for Electro-Optical Ultrasound Data Acquisition}

\author{
\IEEEauthorblockN{
    Marziyeh~Rezaei\textsuperscript{1},
    Sridhar~Singhal\textsuperscript{1},
    Alp A.~Derin\textsuperscript{2},
    Soner~Sonmezoglu\textsuperscript{2},
and~Sajjad~Moazeni\textsuperscript{1}}

\IEEEauthorblockA{\textsuperscript{1}Department of Electrical and Computer Engineering, University of Washington, Seattle, WA}

\IEEEauthorblockA{\textsuperscript{2}Department of Electrical and Computer Engineering, Northeastern University, Boston, MA}
        }
        
\maketitle

%\IEEEauthorblockN{
%    Marziyeh~Rezaei,~\IEEEmembership{Student,~IEEE}
%    Sridhar~Singhal,~\IEEEmembership{Student,~IEEE}
%    Soner~Sonmezoglu,~\IEEEmembership{Member,~IEEE}
%and~Sajjad~Moazeni,~\IEEEmembership{Member,~IEEE}}
\begin{abstract}
This paper presents the design and experimental characterization of a 32-channel, WDM-compatible electronic-photonic receiver array fabricated in the GlobalFoundries 45nm monolithic silicon photonics process. Optimized for miniaturized, millimeter-scale electro-optical ultrasound data acquisition probes for broadband photo-acoustic imaging (PAI), the chip utilizes integrated micro-ring resonator (MRR) demultiplexers and photodiode (PD) arrays to decode multi-wavelength acoustic signals outside the body. This work focuses specifically on the comprehensive characterization of the receiver front-end, evaluating both the optical WDM filter bank and the transimpedance amplifier (TIA) performance. Monolithic co-integration minimizes critical photodiode-to-TIA interface parasitics. The TIA incorporates a DC-offset cancellation loop that compensates for photodiode-induced offset currents while providing tunable low-frequency cutoff control over a range of 30 kHz to 1.5 MHz. This tunability enables effective suppression of flicker noise and improves overall signal integrity. The front-end achieves the peak transimpedance gain of 144 dB$\Omega$ while maintaining an ultra-low input-referred noise current density of $1\text{ pA}/\sqrt{\text{Hz}}$, demonstrating a highly scalable architecture for high-sensitivity intravascular and endobronchial imaging.
\end{abstract}

\begin{IEEEkeywords}
TIA, Silicon Photonics, Ultra-sound, Data Acquisition, Endoscope 
\end{IEEEkeywords}
\section{Introduction}
\label{sec:Introduction}

%Paragraph 1: The Clinical Motivation & The Cable Bottleneck
%In vivo photoacoustic and high-frequency ultrasound imaging are critical modalities for diagnosing early-stage malignancies, such as lung cancer, and assessing cardiovascular disease~\cite{PAI-review}. Applications such as intravascular ultrasound (IVUS) and endobronchial bronchoscopy demand aggressively miniaturized, millimeter-scale probe tips capable of high-resolution 3D imaging. However, traditional acousto-electrical transducer arrays suffer from a fundamental packaging bottleneck: routing high-frequency electrical signals from dense 2D sensor arrays requires bulky coaxial cables. These cables compromise probe flexibility and introduce substantial parasitic capacitance (on the order of hundreds of picofarads) that dramatically degrades the SNR. While integrating active receiver ICs directly inside the probe tip can mitigate cable noise~\cite{JSSC-PAICMUT}, the resulting power dissipation causes dangerous local tissue heating, presenting a severe barrier to clinical adoption.

In vivo photoacoustic and high-frequency ultrasound imaging are important modalities for diagnosing early-stage malignancies, such as lung cancer, and assessing cardiovascular disease~\cite{PAI-review,Dooms169-177,Lin2022TheER}. Applications such as intravascular ultrasound (IVUS) and endobronchial bronchoscopy require millimeter-scale probe tips capable of high-resolution imaging while maintaining flexibility, low power dissipation, and scalable channel count. However, traditional acousto-electrical transducer arrays suffer from a fundamental packaging bottleneck: routing high-frequency electrical signals from dense 2D sensor arrays requires bulky coaxial cables~\cite{cable-count-PAI}. These cables compromise probe flexibility and introduce substantial parasitic capacitance, on the order of hundreds of picofarads, that can significantly degrade the signal-to-noise ratio (SNR). While integrating active receiver ICs directly inside the probe tip can mitigate cable noise~\cite{JSSC_2017_28nm}, the resulting power dissipation can lead to local tissue heating concerns, presenting a major challenge for clinical translation.

%Paragraph 2: The Acousto-Optic Architecture & Circuit Challenges
%To circumvent these constraints, we propose a new paradigm using optical read-out via fibers as illustrated in Fig.~\ref{fig:concept}. In this approach the acoustic waves are converted directly into the optical domain at the probe tip using piezoelectric micromachined ultrasonic transducers (pMUTs)~\cite{Cai2022PhotoacousticIB} coupled to optical microresonators (OMR)~\cite{Soner-OMR}. By encoding the acoustic data into optical signals via wavelength-division multiplexing (WDM), hundreds of sensor channels can be simultaneously interrogated using a single optical fiber, dropping the electrical power consumption at the probe tip to near-zero. However, this shifts the burden of signal integrity to an external electronic-photonic ultrasound (US) data acquisition (DAQ) module. Given the relative weak US to optical modulation efficiencies, the primary circuit bottleneck becomes the receiver tansimpedance amplifier (TIA) front-end, which requires highly sensitive, multi-channel optical demultiplexing and low-noise optical readouts. Designing a scalable array of TIAs for these optical channels demands strict optimization to resolve the fundamental trade-offs between an ultra-low noise floor, adequate ultrasound bandwidth, and variable dynamic range and bandwidth.

\begin{figure}[tp]
\noindent
  \centering{\includegraphics[width=\linewidth]{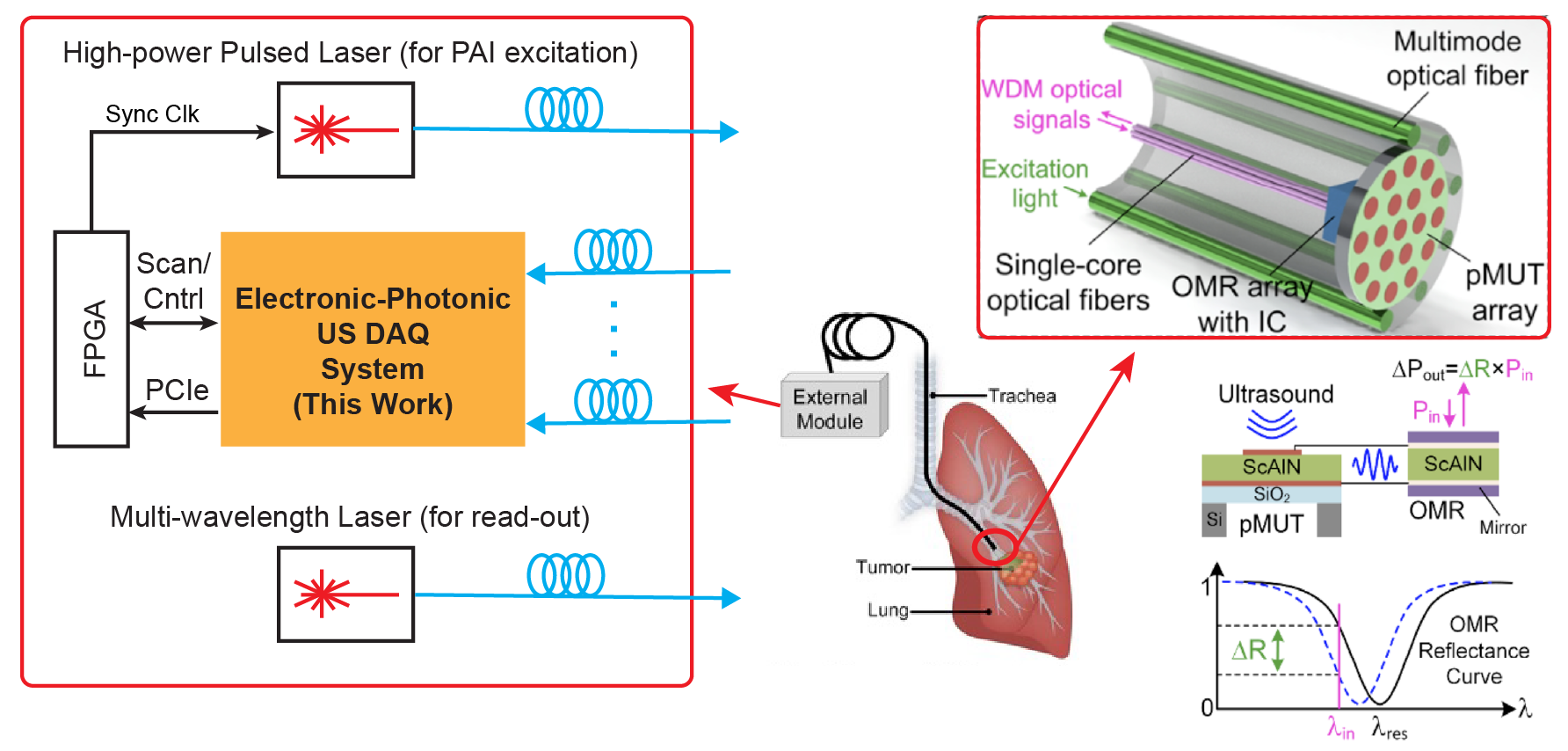}}
\caption{Proposed electro-optical endoscope/bronchoscope concept for PAI using direct US-to-optical transduction on the probe tip.}
\label{fig:concept}
\vspace{-0.6cm}
\end{figure}

To circumvent these constraints, we propose an optical readout approach using fibers, as illustrated in Fig.~\ref{fig:concept}. In this approach, acoustic waves are converted directly into the optical domain at the probe tip using piezoelectric micromachined ultrasonic transducers (pMUTs)~\cite{Cai2022PhotoacousticIB} coupled to optical microresonators (OMRs)~\cite{Soner-OMR}. By encoding acoustic data into optical signals via wavelength-division multiplexing (WDM), many sensor channels can be interrogated through a single optical fiber, enabling near-zero electrical power dissipation at the probe tip. However, this architecture shifts the signal-integrity burden to an external electronic-photonic ultrasound (US) data-acquisition (DAQ) module. Given the relatively weak ultrasound-to-optical modulation efficiency, the primary circuit bottleneck becomes the receiver TIA front-end, which must provide highly sensitive, multi-channel optical demultiplexing and low-noise optical readout. Designing a scalable array of TIAs for these optical channels requires careful optimization of the trade-offs among ultra-low noise floor, sufficient ultrasound bandwidth, dynamic range, and channel scalability.

%Paragraph 3: This Work & Concrete Contributions
This paper addresses these analog front-end (AFE) challenges by presenting the implementation and comprehensive circuit characterization of a 32-channel WDM-compatible electronic-photonic ultrasound receiver front-end fabricated in a commercial GlobalFoundries 45nm monolithic silicon photonics process (GF45SPCLO)~\cite{rakowski202045nm}. Unlike prior demonstrations that focused on single-channel optical ultrasound readout, discrete photonic components, or limited receiver integration, this work combines 32-channel WDM-compatible reception, monolithic CMOS-photonic integration, compact DWDM demultiplexing, reduced PD-TIA parasitics, and programmable TIA bandwidth in a single scalable receiver platform. Our focus in this work is characterizing the core components of the receiver front-end, including the optical dense WDM (DWDM) filter bank and the TIA. This architecture utilizes an array of double micro-ring resonators (MRRs) for ultra-compact optical demultiplexing. Each channel incorporates a programmable tuning mechanism that adjusts the TIA low-cutoff frequency from 30 kHz to 1.5 MHz, enabling flicker-noise mitigation. The upper-cutoff frequency can be configured to either 30 MHz or 15 MHz, depending on the application requirements. Experimental characterization of the front-end demonstrates WDM filtering performance, an overall peak transimpedance gain of 144 dB$\Omega$, and an ultra-low input-referred noise current density of 1 $pA / \sqrt{Hz}$, establishing the receiver front-end as a scalable electronic-photonic DAQ architecture for fiber-based ultrasound/PAI systems.

%This paper addresses these analog front-end (AFE) challenges by presenting the implementation and comprehensive circuit characterization of a 32-channel WDM-compatible electronic-photonic US receiver front-end fabricated in a commercial GlobalFoundries 45nm monolithic silicon photonics process (GF45SPCLO). Our focus in this work is dedicated specifically to characterizing the core components of the receiver front-end, detailing the performance of the optical dense WDM (DWDM) filter bank alongside the individual TIA channel metrics. By integrating advanced 45nm CMOS circuitry monolithically with silicon photonic components on a single die, this architecture utilizes an array of double micro-ring resonators (MRRs) for ultra-compact optical demultiplexing while minimizing the critical parasitic capacitance at the photodiode-TIA interface. Each channel incorporates a programmable tuning mechanism that dynamically adjusts the TIA low-cutoff frequency from 10 kHz to 1.5 MHz, enabling flicker-noise mitigation. The upper-cutoff frequency can be configured to either 30 MHz or 15 MHz, depending on the application requirements. Experimental characterization of the front-end demonstrates excellent WDM filtering performance, a peak trans-impedance gain of 144 dBohm, and an ultra-low input-referred noise current density of 5pA/Hz, proving the scalability of monolithic electronic-photonic architectures for next-generation biomedical DAQ systems.
%\input{architecture}

\section{Electro-optical US DAQ System }
\label{sec:usdaq}

\begin{figure}[tp]
  \centering{\includegraphics[width=80mm]{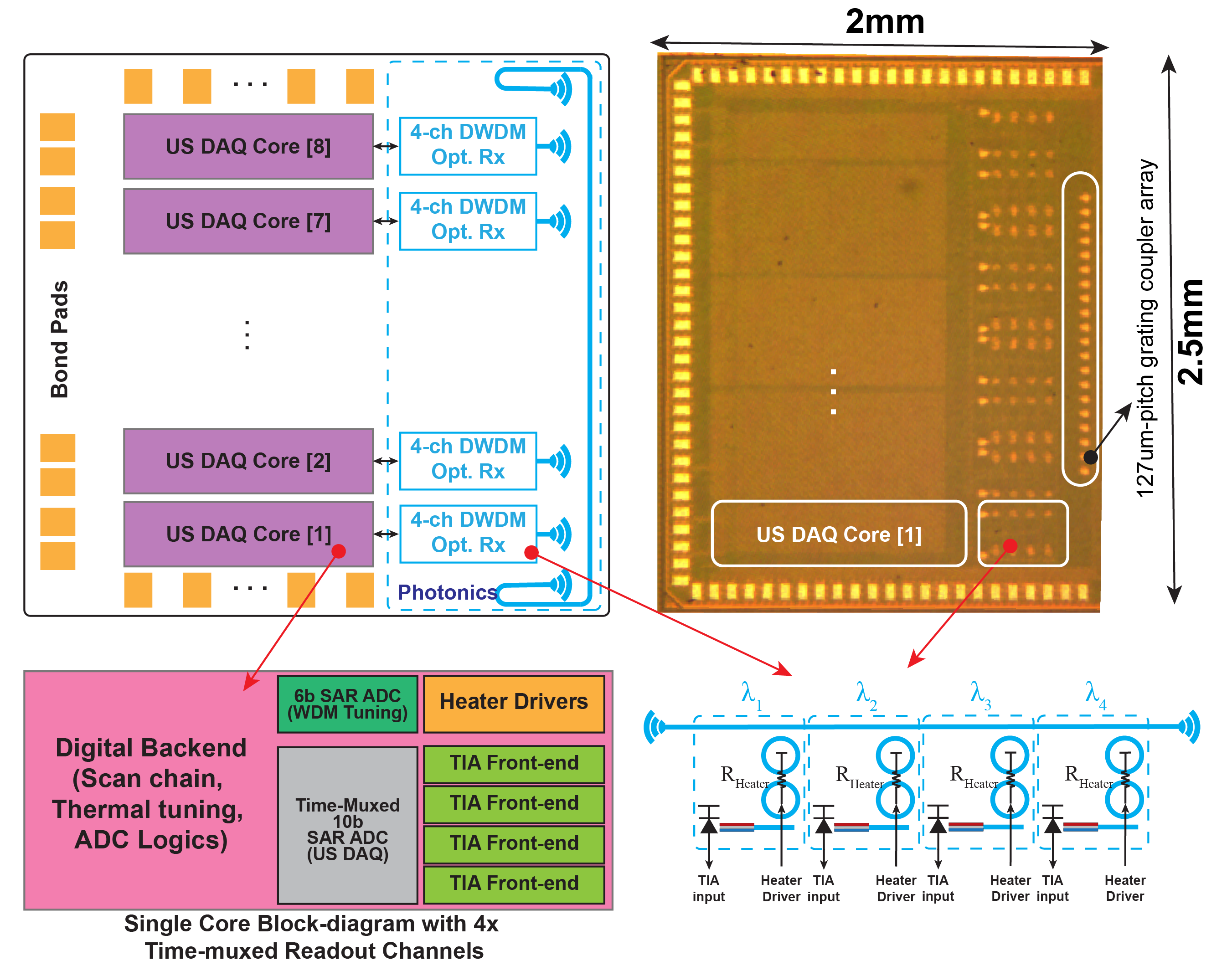}}
\caption{Chip-level block-diagram of electro-optical US-DAQ system in GF45SPCLO.}
\label{fig:chip_arch}
\vspace{-0.6cm}
\end{figure}

Fig.~\ref{fig:chip_arch} illustrates the proposed US DAQ architecture. The system interfaces with eight input optical fibers (using grating coupler and a fiber array), each carrying four wavelength channels, for a total of 32 optical channels. Each "core" is a group of four TIA AFEs, followed by a shared time-multiplexed 10-b SAR ADC (shared for 4-WDM channels). The digitized output is then fed into a digital backend and transmitted to an FPGA via PCIe 3.0 using LVDS signaling (The ADC characterization and data link to FPGA is not discussed in this paper). Each core, also includes a 6-bit SAR ADC to read-out the optical power of each channel, and use a thermal tuning digital loop~\cite{moazeni-jssc2017} to tune the heater strength in each MRR to lock the wavelength of each optical filter.

Within each core, an optical demultiplexer based on double-ring MRR filters separates the wavelength channels and routes each channel to a corresponding PD. The PDs convert the modulated optical signals into electrical photocurrents. Each MRR has a flat-top optical bandwidth of 1 nm (for 1 dB bandwidth) with $\sim$0.25 dB insertion loss (IL) in the O-band (1310 nm). To compensate for process variations, thermal drift, and wavelength-spacing errors, each MRR is equipped with an integrated heater. The heaters are driven by pulse density modulation (PDM), whose control words are programmed through a scan chain in the digital backend. While the channel spacing is tunable, our initial nominal design is for 1 nm channel spacing to match our DWDM laser source. 

The resulting photocurrents are amplified by the TIA AFE. Designing such TIAs presents several challenges due to the trade-offs among gain, bandwidth, and noise. Because the transduced signal can be relatively weak, the TIA must tolerate DC offsets that may be up to 100× larger than the signal of interest. This requires dynamic DC cancellation, which can itself contribute significantly to the overall AFE noise budget, motivating the use of noise-optimization techniques.

\begin{figure*}[htbp]
\centering{\includegraphics[width=140mm]{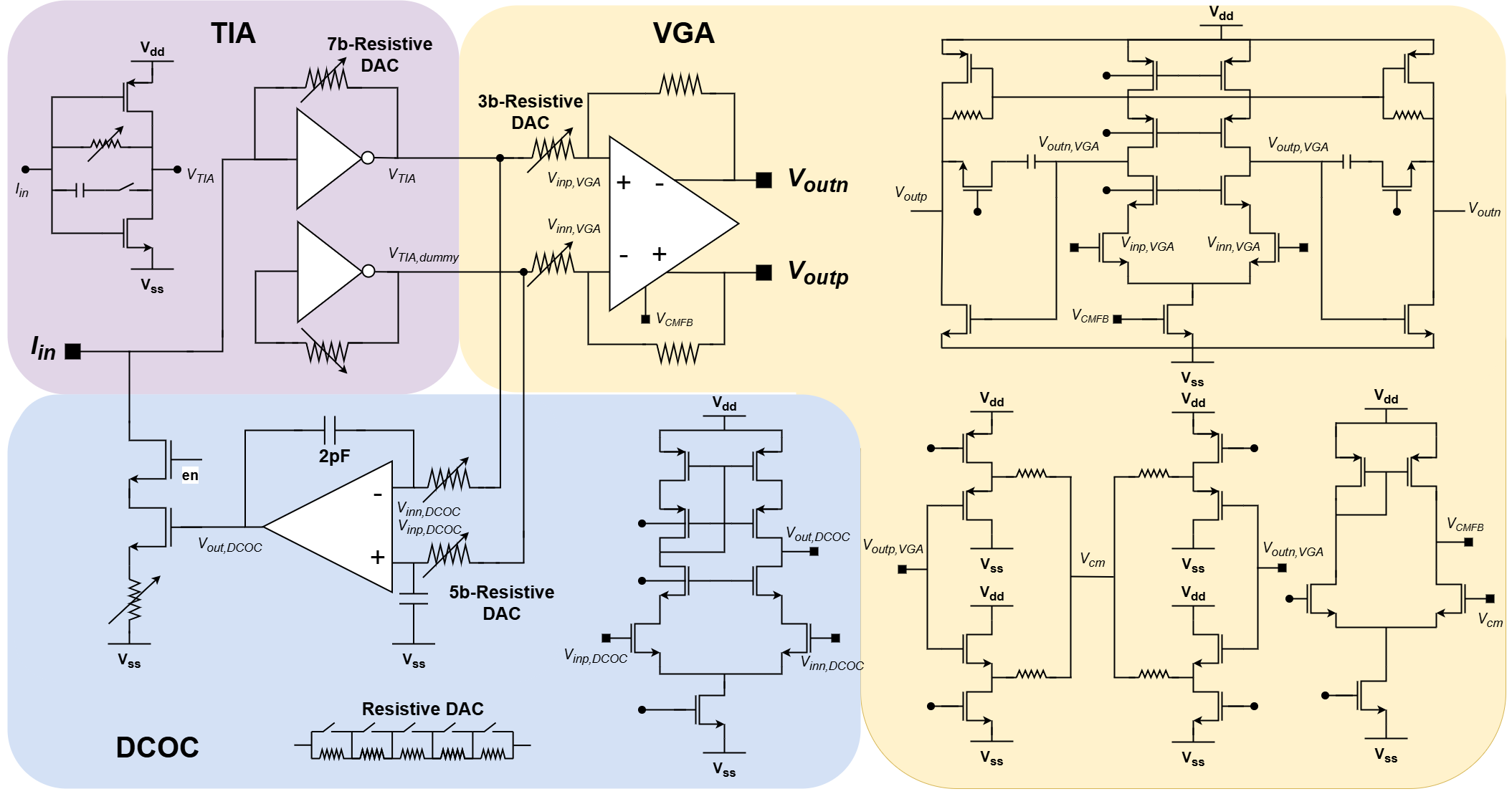}}
\caption{Circuit block-diagram of the ultra-low noise TIA front-end}
\label{fig:block diagram}
\vspace{-0.5cm}
\end{figure*}

As shown in Fig.~\ref{fig:block diagram}, each AFE includes a tunable-gain TIA, a DC-offset cancellation (DCOC) loop with programmable low-cutoff frequency, and a variable-gain amplifier (VGA). This programmable gain chain enables the receiver to accommodate the wide dynamic range (65 dB) associated with variable optical attenuation and tissue penetration depth.

The TIA is implemented using an inverter-based feedback architecture. A 7-bit resistive feedback DAC programs the transimpedance gain, enabling the receiver sensitivity to be adjusted according to the input signal level. The TIA output is then fed to a DCOC loop, which extracts the DC and low-frequency components using an active low-pass filter implemented with a telescopic operational amplifier in an RC-feedback configuration. The extracted offset is subsequently canceled to suppress baseline drift. In addition to offset cancellation, the loop attenuates low-frequency noise components, including flicker noise, which can significantly contribute to the noise floor in low-bandwidth systems. 

The extracted offset is converted into a compensation current through a transconductance (Gm) stage and subtracted from the front-end signal path. The Gm stage is realized using an NMOS transistor with source degeneration. Source degeneration reduces the thermal noise contribution of the Gm device and the opamp noise in the cancellation path, which is one of the dominant contributors to the front-end noise in this system. The degeneration element is implemented using an NMOS transistor operating in the triode region, whose gate voltage is programmable to provide two selectable DC-offset cancellation range of 0-30 $\mu$A , and 31-60 $\mu$A. A 5-bit resistive DAC in the DCOC path further tunes the low-cutoff frequency, enabling flicker-noise attenuation.

Following TIA and offset cancellation, the differential VGA provides additional programmable gain. The VGA is implemented using a two-stage operational amplifier in a resistive-feedback configuration, with its gain controlled through a 3-bit resistive-feedback DAC allowing the AFE gain from 79 dB$\Omega$ to 144 dB$\Omega$. Finally, the VGA output is connected to a buffer stage to support probing and characterization.

\section{Measurement Results}
\label{sec:results}

The die photograph of the US DAQ chip is shown in Fig.~\ref{fig:chip_arch}, with 2 mm × 2.5 mm dimensions. The US DAQ chip is wire-bonded to a PCB for testing and characterization. The chip operates with two supply domains: 1.8 V for the AFE and PD biasings, and 1.0 V for the digital backend. All configuration bits are programmed through the scan chain via an FPGA.

\begin{figure}
\centering{\includegraphics[width=70mm]{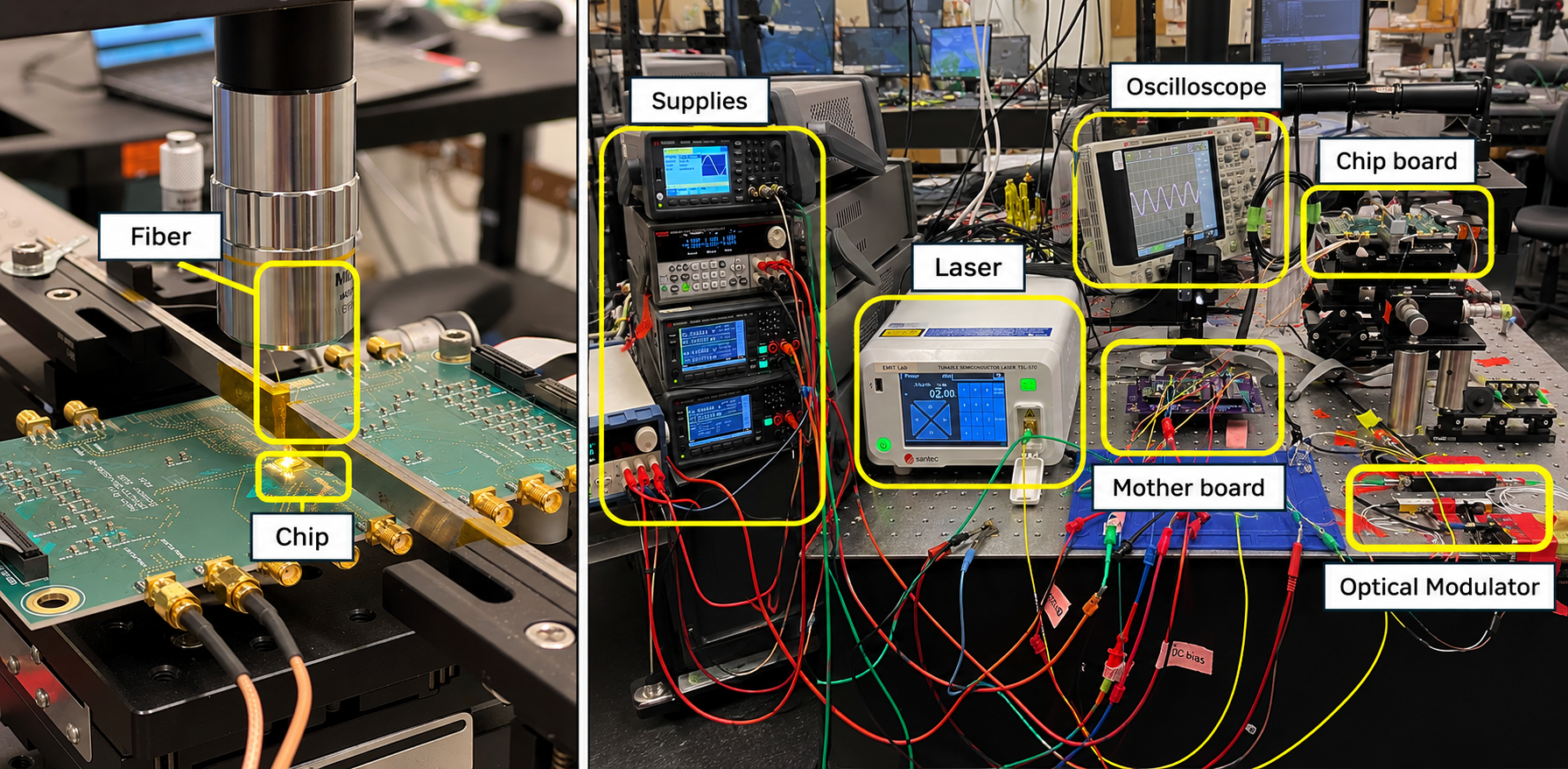}}
\caption{Photo of the test setup (the US signal is emulated using off-chip commercial optical modulator).}
\label{fig:setup}
\vspace{-0.6cm}
\end{figure}

\begin{figure}
\centering{\includegraphics[width=80mm]{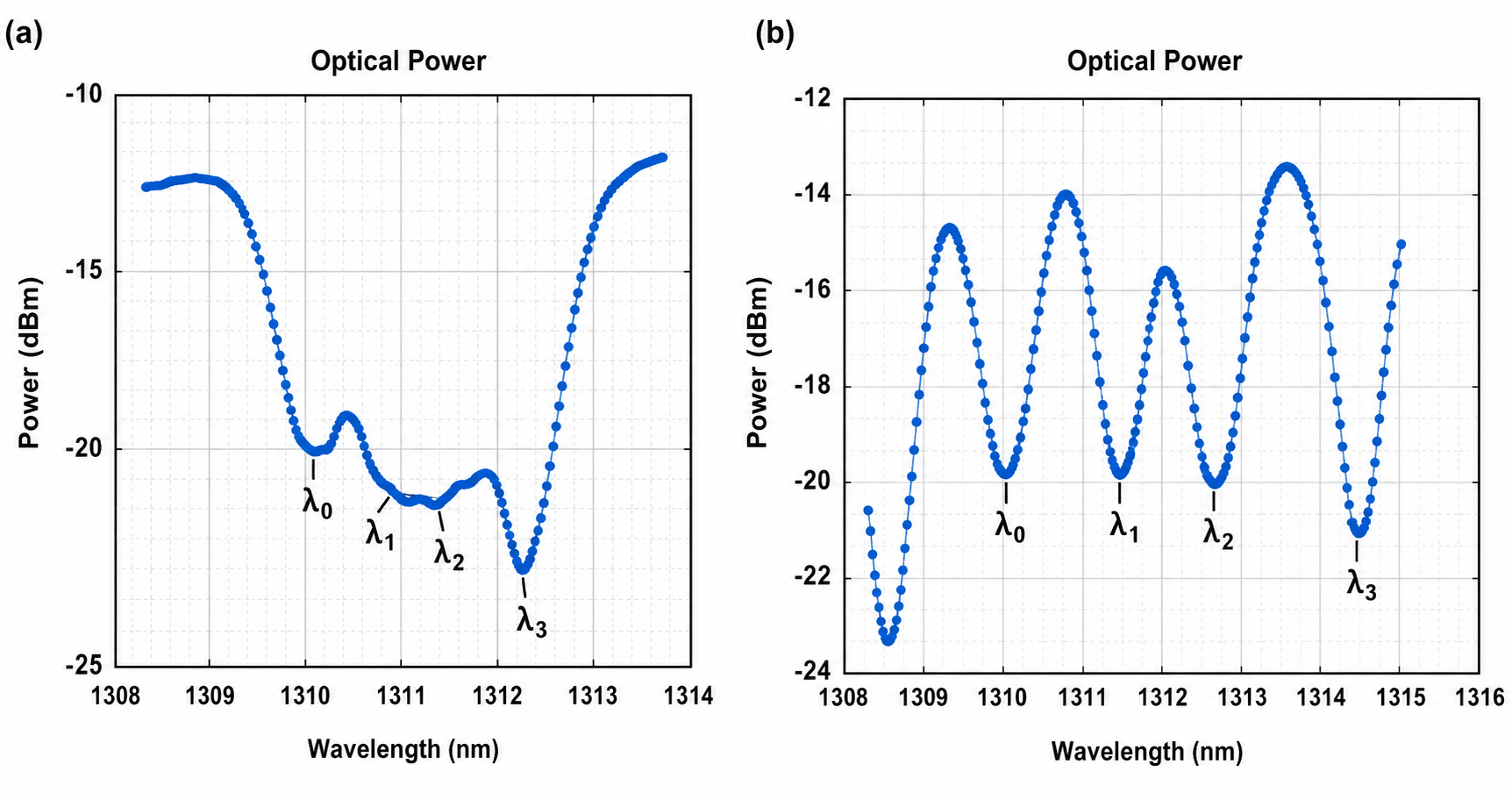}}
\caption{Measured optical response (through-port) of a single DWDM DeMUX channel before (a) and after (b) the thermal tuning loop corrects the channel spacings.}
\label{fig:WDM_spectrum}
\vspace{-0.6cm}
\end{figure}

Fig.~\ref{fig:setup} shows the experimental setup in the lab. First, we characterized the optical MRRs using external tunable laser (at 13 dBm power and fiber-to-chip coupling loss of ~6.5dB). The laser wavelength was then swept to identify the resonance wavelengths of the MRRs using a through-port grating coupler at the output of the bus waveguide. As shown in Fig.~\ref{fig:WDM_spectrum} (a), there are four resonances but in this example two resonance wavelengths, $\lambda_{1}$ and $\lambda_{2}$, are closely spaced (due to process variation), resulting in a flattened spectral response in that region. Next, thermal tuning was used to adjust the resonance wavelengths and achieve the desired channel spacing. The PDM control codes were programmed to the target heater settings, allowing the PDM driver to regulate the heater power and shift the MRR resonances accordingly. Fig.~\ref{fig:WDM_spectrum} (b) shows the measured output spectrum from the bus waveguide after thermal tuning, where the resonance wavelengths were successfully adjusted to a channel spacing of 1.2 nm.

\begin{figure}
\centering{\includegraphics[width=\linewidth]{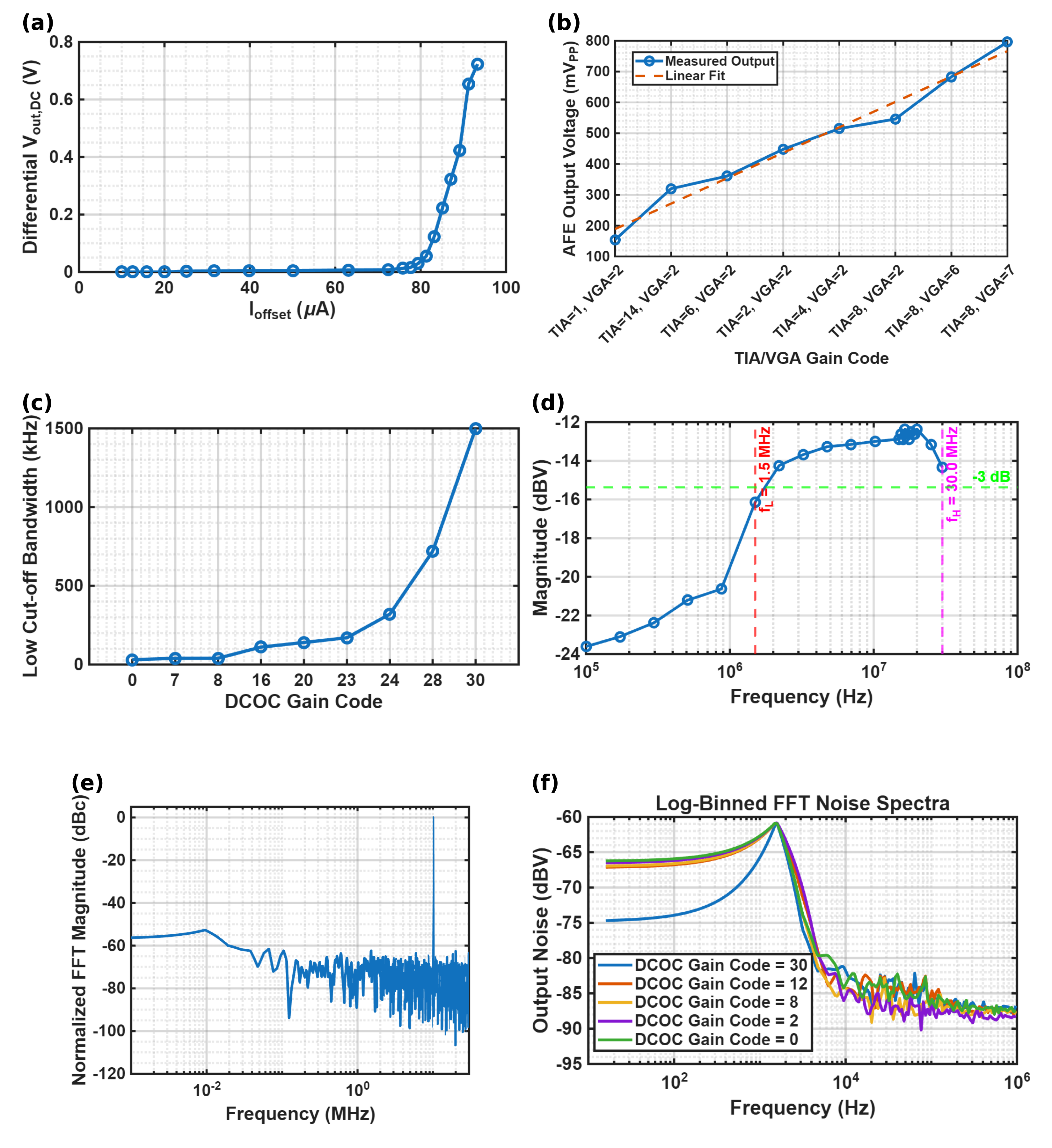}}
\caption{TIA measurement results: (a) DCOC offset cancellation range; (b) AFE gain linearity; (c) DCOC low-cutoff bandwidth tunability; (d) measured AC response; (e) normalized measured AFE output FFT for a 10MHz input; (f) DCOC low-frequency noise shaping versus different gain code.}
\label{fig:result}
\vspace{-0.6cm}
\end{figure}

Next, the laser wavelength was set to the resonance wavelength of the MRR connected to the AFE channel under test. The DC-offset cancellation functionality was first characterized by sweeping the laser power from -10 dBm to 10 dBm, resulting in a photocurrent-induced DC offset ranging from 0 to 100 $\mu$A at the TIA input. The measured differential DC output of the AFE is shown in Fig.~\ref{fig:result} (a). As shown in here, the DC-offset cancellation loop successfully compensates for input DC offsets of up to 65 $\mu$A without noticeable error. For larger offsets, the loop begins to lose lock, which is consistent with the design specifications.

Next, the linearity of the AFE gain was characterized with respect to a constant input level. For this measurement, the laser output was externally modulated with a 10 MHz sinusoidal signal and coupled into the bus waveguide, generating an input current of 3 $\mu$A at the TIA input. The TIA gain code was then swept to vary the gain, and the corresponding output amplitude was measured. Fig.~\ref{fig:result} (b) shows the measured TIA gain as a function of the programmed gain code. The measured response exhibits linearity, with an average mean-squared-error (MSE) of 6.9$\%$ relative to the fitted linear trend.

As discussed in the previous section, the DCOC enables tuning of the AFE low-cutoff frequency through its gain code. This tuning is achieved by adjusting the effective resistance in the feedback path, which in turn shifts the location of the DCOC pole. Fig.~\ref{fig:result} (c) shows the measured low-cutoff frequency as a function of the DCOC gain code. As shown, the DCOC supports a tunable low-cutoff frequency range from 30 kHz to 1.5 MHz. It is worth noting that this tuning range can be further extended by increasing the capacitance in the feedback path.

Fig.~\ref{fig:result} (d) shows the measured AC frequency response of the system obtained by sweeping the input frequency. The measured upper-cutoff frequency is 30 MHz. By enabling the feedback capacitor in the TIA feedback path, the upper-cutoff frequency can be reduced to 16 MHz, allowing the bandwidth to be adjusted according to the application requirements. The characterization was limited to 30 MHz due to the maximum output frequency supported by the waveform generator.

Fig.~\ref{fig:result} (e) presents the normalized FFT of the AFE output for a 10 MHz with 1 $\mu$A amplitude at the TIA input. The measured signal-to-noise-and-distortion ratio (SNDR) is obtained by integrating the noise over 5 MHz bandwidth which results in 50.7 dB SNDR and 1 pA/$\sqrt{\mathrm{Hz}}$ input-referred noise (AFE gain is 114 dB$\Omega$).

Next, the flicker-noise reduction performance of the AFE was evaluated. A key distinction of this work compared to prior implementations is the use of a feedback-based DCOC loop, which suppresses low-frequency components, including flicker noise, by continuously canceling the DC and near-DC offsets at the input. As shown in Fig.~\ref{fig:result} (f), configuring the DCOC cutoff frequency to 1.5 MHz reduces the low-frequency noise floor by approximately 10 dB compared to the lower-cutoff setting.

%%%%%%% Table %%%%%%%%%%
\begin{table*}[!t]
\centering
\caption{Comparison with state-of-the-art works.}
\begin{tabular}{|c|c|c|c|c|c|c|c|}
\hline
\textbf{Parameter} &
{\cellcolor{yellow}\textbf{This work}} &
\textbf{JSSC 2023~\cite{JSSC_2023_45nm}} &
\textbf{JSSC 2023}~\cite{JSSC_2023_180nm} &
\textbf{ISSCC 2021~\cite{ISSCC_2021}} &
\textbf{JSSC 2017~\cite{JSSC_2017_180nm}} &
\textbf{JSSC 2017 \cite{JSSC_2017_28nm}}\\
\hline
Technology &
{\cellcolor{yellow}45 nm SOI } &
45\,nm SOI &
180\,nm BCD &
130\,nm BCD &
180\,nm &
28\,nm SOI \\
\hline

Sensing Technique &
{\cellcolor{yellow}Electro-Optical } &
Electro-Optical &
Electrical &
Electrical &
Electrical &
Electrical \\
\hline

Number of Channels &
{\cellcolor{yellow}} 32&
32 &
64 &
8960 &
864 &
16 \\
\hline

AFE Bandwidth (MHz) &
{\cellcolor{yellow}30} &
N/R &
8 &
10 &
6 &
10 \\
\hline
Dynamic Range (dB) &
{\cellcolor{yellow}} 65&
74 &
82 &
15 &
48 &
90 \\
\hline
SNDR (dB) &
{\cellcolor{yellow} 50.7 } &
25.5 &
54 &
54.1 &
51 &
41.9 \\

\hline
Closed-loop DC-Offset Cancellation&
{\cellcolor{yellow} Yes} &
No &
No (AC coupled) &
No &
No (AC coupled)&
No \\

\hline
Low-cutoff Bandwidth (MHz)&
{\cellcolor{yellow}} 0.03 - 1.5&
N/R &
N/R &
N/R &
0.3&
N/R \\

\hline

Power/Channel (mW) &
{\cellcolor{yellow}3.94} &
2.63 &
1.17 &
0.43 &
0.27 &
17.5 \\
\hline
Active Area/Channel (mm$^2$) &
{\cellcolor{yellow}0.014} &
0.034 &
0.04 &
0.01 &
0.0225 &
0.041 \\
\hline
\multicolumn{7}{l}{\footnotesize N/R: Not Reported} \\
\end{tabular}
\label{tab:comparison}
\vspace{-1cm}
\end{table*}
\section{Conclusion}
\label{sec:conclusion}
This work presents a 32-channel WDM-compatible electronic-photonic ultrasound receiver front-end fabricated in the GlobalFoundries 45 nm monolithic silicon photonics process. The AFE integrates a programmable TIA and a tunable DC-offset cancellation loop for flicker-noise suppression.

While only one channel result is presented, due to the limited pad availability for externally probing the AFE outputs, the remaining channels have been optically verified to exhibit the same transmission function. The electronics of these channels are currently being further tested using the on-chip ADC.

Table.~\ref{tab:comparison} summarizes the results and compares our work with previous works on high-channel US DAQ systems. In addition to the power overhead associated with integrating electrical ICs at the probe tip, purely electrical solutions suffer from a strong trade-off between SNDR and BW due to excess thermal noise and the direct parasitic capacitance of p/cMUTs. Compared with the only previously reported electro-optical work~\cite{JSSC_2023_45nm}, our design employs dynamic DC current cancellation instead of an external current DAC. Furthermore, the proposed noise-suppression techniques and closed-loop DCOC architecture enable higher SNDR and BW. These results validate monolithic electronic-photonic receiver arrays as a promising solution for miniaturized PAI systems. Future work will focus on demonstrating end-to-end system performance using our pMUT and OMR technologies.

\section*{Acknowledgment}

%This work is supported by ARPA-H (PAIL: PhotoAcoustic Imaging technology for diagnostic Lung assessment). 

Research reported in this publication was supported by the Advanced Research Projects Agency for Health (ARPA-H) under Award Number D24AC00380-00. The content is solely the responsibility of the authors and does not necessarily represent the official views of ARPA-H. The authors would also like to thank GlobalFoundries for providing silicon fabrication through the 45SPCLO university partnership program.

\FloatBarrier
\bibliographystyle{IEEEtran}
\bibliography{reference}
\end{document}